# Spin-Hall nano-oscillator with oblique magnetization and Dzyaloshinskii-Moriya interaction as generator of skyrmions and nonreciprocal spin-waves

A. Giordano[1], R. Verba[2], R. Zivieri[3], A. Laudani[4], V. Puliafito[5], G. Gubbiotti[6], R. Tomasello[7], G. Siracusano[1], B. Azzerboni[5], M. Carpentieri[8], A. Slavin[9], and G. Finocchio[1*]

[1] Department of Mathematical and Computer Sciences, Physical Sciences and Earth Sciences, University of Messina, Messina, Italy

[2] Institute of Magnetism, National Academy of Sciences of Ukraine, Kyiv, Ukraine

[3] Department of Physics and Earth Sciences and CNISM Unit of Ferrara, University of Ferrara, Ferrara, Italy

[4] Department of Engineering, University of Roma Tre, Roma, Italy

[5] Department of Engineering, University of Messina, Messina, Italy

[6] Istituto Officina dei Materiali del CNR (CNR-IOM), Sede Secondaria di Perugia, c/o Dipartimento di Fisica e Geologia, University of Perugia, Perugia, Italy

[7] Department of Engineering, Polo Scientifico Didattico di Terni, University of Perugia, Terni, Italy

[8] Department of Electrical and Information Engineering, Politecnico di Bari, I-70125 Bari, Italy

[9] Department of Physics, Oakland University, Rochester, MI 48309, USA

**Abstract**

Spin-Hall oscillators (SHO) are promising sources of spin-wave signals for magnonics applications, and can serve as building blocks for magnonic logic in ultralow power computation devices. Thin magnetic layers used as "free" layers in SHO are in contact with heavy metals having large spin-orbital interaction, and, therefore, could be subject to the spin-Hall effect (SHE) and the interfacial Dzyaloshinskii-Moriya interaction (*i*-DMI), which may lead to the nonreciprocity of the excited spin waves and other unusual effects. Here, we analytically and micromagnetically study magnetization dynamics excited in an SHO with oblique magnetization when the SHE and *i*-DMI act simultaneously. Our key results are: (i) excitation of nonreciprocal spin-waves propagating perpendicularly to the in-plane projection of the static magnetization; (ii) skyrmions generation by pure spin-current; (iii) excitation of a new spin-wave mode with a spiral spatial profile originating from a gyrotropic rotation of a dynamical skyrmion. These results demonstrate that SHOs can be used as generators of magnetic skyrmions and different types of propagating spin-waves for magnetic data storage and signal processing applications.

# Introduction

Spin-orbitronics combined with other sub-fields of spintronics, such as magnonics and spin-caloritronics, has created a novel paradigm in information processing which could become a viable alternative to CMOS electronics[1].

Recent experimental and theoretical developments in spin-orbitronics have clearly shown a great potential in generation of spin-currents able to compensate damping in magnetic materials[2,3,4,5,6]. The spin-Hall effect (SHE) plays a dominant role in the above-mentioned experiments, as it converts the input *charge current*, flowing in a heavy metal, into a *spin-current*, diffusing perpendicularly into the adjacent ferromagnet, and creating a spin-transfer torque (STT) that acts on the ferromagnet magnetization[7]. Another interesting and highly non-trivial spin-orbital effect is the interfacial Dzyaloshinskii-Moriya interaction (*i*-DMI)[8]. Both SHE and *i*-DMI have been used to improve the performance of "racetrack" device prototypes in magnetic storage[9,10], to add a new degree of freedom in the design of magnetoresistive memories[3,11], to create nonreciprocity in the spin-wave propagation for signal processing applications[12,13,14], to excite coherent magnetization self-oscillations[4,5], and for the manipulation of skyrmions in ultrathin ferromagnetic materials[15,16]. However, to the best of our knowledge, the influence of *i*-DMI on the performance of a spin-Hall oscillator (SHO) has not been studied so far[4,5].

Here, we present the magnetization dynamics induced by the SHE in a realistic SHO structure, taking into account the influence of the *i*-DMI[8]. We have chosen a state-of-the-art SHO geometry (Fig.1a) where the charge current $I$ flows in the Pt layer along the $x$-axis between the golden electrodes and, due to the SHE in Pt, a spin-current is locally injected into the ultrathin extended CoFe ferromagnet (SHO "free" layer). The CoFe layer has an in-plane easy axis at zero bias field, so when a sufficiently large out-of-plane bias field is applied at an oblique angle in the "$yz$" plane (Fig.1b), the static magnetization **M** of the "free" layer also goes out-of-plane, making the angle $\theta_M$ with the vertical axis "$z$". In such a geometry, the Slonczewski *propagating* spin waves[17] can be excited in any in-plane direction[18,19,20] and, due to the influence of the *i*-DMI, they have the maximum nonreciprocity when propagating along the $x$-axis, perpendicular to the in-plane projection of the bias magnetic field. Our numerical simulations have shown that the wave numbers of spin waves excited at a particular frequency $\omega$ and propagating along the positive and negative directions of the $x$-axis are different. The difference is proportional to the magnitude of the *i*-DMI parameter $D$. This result, well reproduced by a simple one-dimensional analytical model, can be used to establish a novel procedure for the experimental measurements of $D$. Micromagnetic simulations have also demonstrated that (*i*) a novel propagating spin-wave mode, characterized by a spiral spatial profile, can be excited at sufficiently large magnitudes of $D$ and $I$, and (*ii*) skyrmions can be efficiently nucleated by the SHE in the SHO geometry (Fig.1c). Similarly to optics [21,22], the excitation of spiral spin-waves in magnetism could be attractive for designing new information coding protocols. Recent experimental observations have demonstrated that skyrmions[15,16,23,24,25] can be nucleated via conversion of domain walls in Ta/CoFeB/MgO[26], or by applying an out-of-plane field in Ir/Co/Pt[27] and Pt/Co/MgO[28] multilayers. Although a single skyrmion can be nucleated by a spin-polarized scanning tunneling microscope[29], the control of its room temperature nucleation is still an experimental challenge. Earlier achievements have shown the possibility to solve this problem[15,23,30]. Our results show an alternative method to control the nucleation of single skyrmions, based on the use of the SHE.

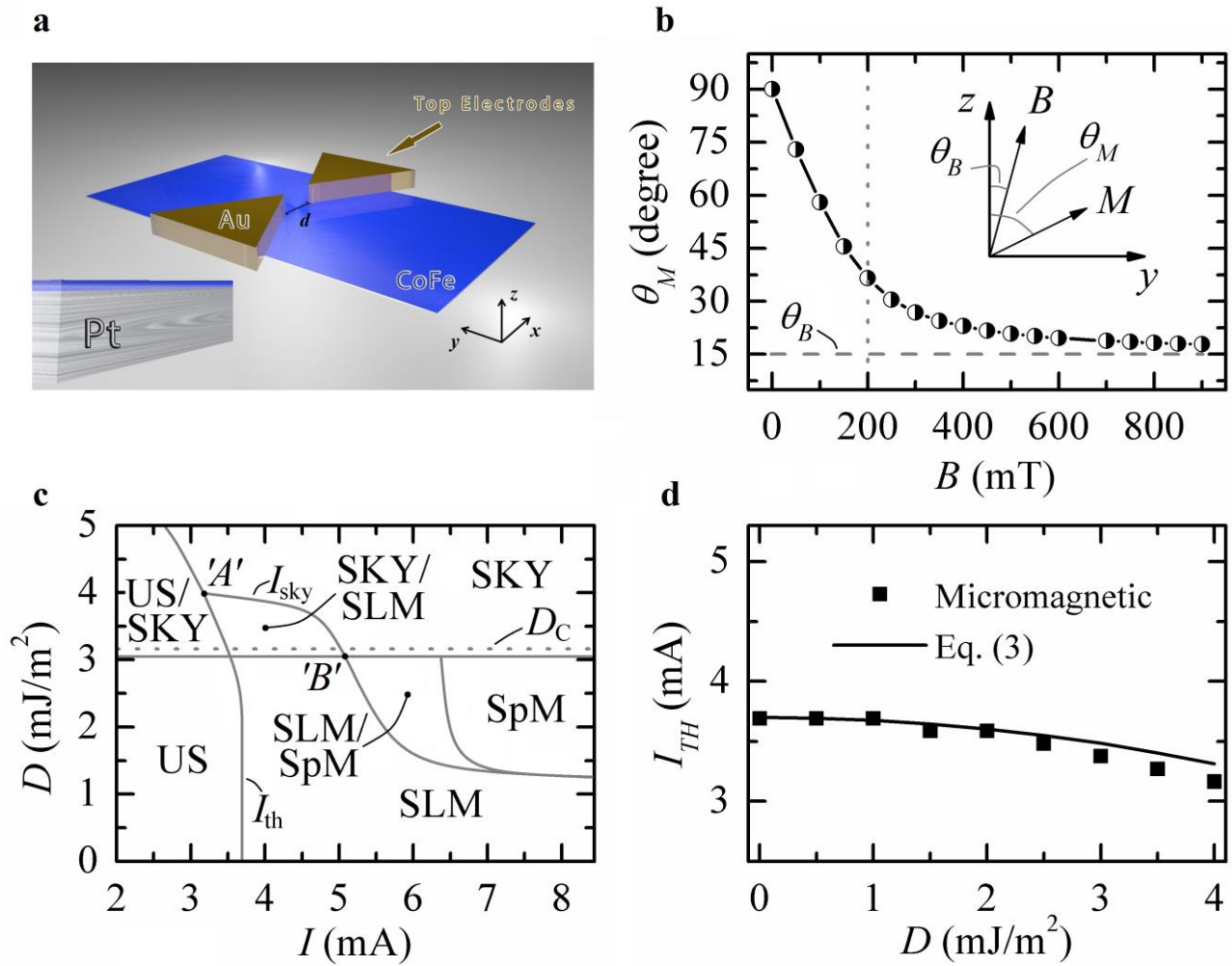

**Figure 1 | Sketch of the SHO device under investigation and dynamical phase diagram of this device. a**, The sketch of a bilayer composed of the CoFe ferromagnetic layer and a layer of a heavy metal (Pt) and having the rectangular cross-section. The thick Au electrodes carry the charge current everywhere, except the inter-electrode gap of the width $d$, where the charge current flows inside the bilayer and excites perpendicular (vertical) spin current going into the CoFe ferromagnetic layer. A rectangular coordinate system for the above described SHO geometry is shown. **b**, The angle $\theta_M$ characterizing the equilibrium direction of the static magnetization in the CoFe ferromagnetic layer as a

function of the magnitude of the external bias magnetic field *B*. The vertical line at *B*=200mT separates the regions where it is possible to excite localized and propagating spin-wave modes, respectively. Inset: Cartesian coordinate reference system where the angles $\theta_M$ and $\theta_B$ are shown explicitly. **c**, The phase diagram of the SHO excitations on the *D* vs. *I* plane. Seven different regions can be distinguished in this phase diagram: uniform states (US), Slonczewski linear modes (SLM), spiral modes (SpM), skyrmions (SKY), uniform states/skyrmions (US/SKY), Slonczewski linear modes/spiral modes (SLM/SpM) and Slonczewski linear modes/skyrmions (SLM/SKY). The amplitude of the external field is *B*=400mT. The Oersted field is included in the model. $D_C$ is the critical value of the *i*-DMI parameter (see explanations below), $I_{th}$ is the threshold current, $I_{sky}$ is the current needed to nucleate skyrmions (line between the points 'A' and 'B'); **d**, Comparison between the threshold current of the SLM excitation obtained by means of micromagnetic simulations (symbols) and using the analytical formula (3) (solid line).

**Results**

**Static characterization of the SHO structure and phase diagram of the SHO excitations.** We have micromagnetically studied a Pt(5nm)/CoFe(1nm) SHO with a rectangular cross section of 1500x3000nm$^2$ (see Fig.1a for the sketch of the device, including a Cartesian coordinate system where *x* and *y* are the in-plane axes, while *z* is the out-of-plane axis, Methods and Supplementary Note 1 for the detailed description of the micromagnetic framework and simulation parameters). Fig.1b shows the angle $\theta_M$, characterizing the equilibrium orientation of the static magnetization in the SHO, as a function of the external bias magnetic field *B*. This field is applied at the tilting angle $\theta_B$=15° with respect to the perpendicular of the SHO ferromagnetic layer in the *y-z* plane (see inset in Fig.1b). As the bias field increases, the magnetization vector tends to align along the field direction.

Similarly to what is observed in STT oscillators based on the point-contact geometry, the type of the spin-wave mode excited by the SHE can be controlled by the direction of the bias magnetic field and the effective anisotropy. In particular, the materials with in-plane easy axis demonstrate excitation of self-localized spin-wave "bullets" for sufficiently large values of $\theta_M$, and excitation of Slonczewski propagating spin-wave modes for sufficiently small values of $\theta_M$[18,31]. In this study, numerical simulations showed that, for the bias field larger than 200mT and $\theta_M$<37°, the Slonczewski propagating spin-wave modes were excited.

As it will be discussed below, the additional degree of freedom of the *i*-DMI can introduce qualitative differences in the spatial profile of the Slonczewski-type cylindrical mode, compared to the case when *i*-DMI is ignored. Hereafter, we focus on the results obtained at the bias field of 400mT and active region (distance between the Au electrodes in Fig.1a) of *d*=100nm, however similar findings have been obtained at *d*=200nm and at larger bias fields (up to 800mT).

Fig. 1c shows a phase diagram of dynamical excitations in the SHO on the plane $D$-vs-$I$. Seven different regions can be identified: (*i*) uniform states (US), (*ii*) Slonczewski linear modes (SLM), (*iii*) spiral modes (SpM), (*iv*) skyrmions (SKY), (*v*) uniform states/skyrmions (US/SKY), (*vi*) Slonczewski linear modes/skyrmions (SLM/SKY) and (*vii*) Slonczewski linear/spiral modes (SLM/SpM). At small values of the driving current, the SHO is in the US, i.e. in a region characterized by a uniform magnetic configuration. SLMs are excited at a critical current $I_{th}$ that slightly decreases as a function of $D$ (see Fig.1d). The excited modes in the SLM region exhibit a two-dimensional radiation pattern that changes from the isotropic (see Supplementary Movies 1 and 2 for the SLM dynamics at $I$=4.22mA and $I$=5.28mA respectively, $B$=400mT and $D$=0.0mJ/m$^2$) to the anisotropic cylindrical profile with the increase of the *i*-DMI parameter $D$ (see Supplementary Movies 3 and 4 for the SLM at $I$=4.22mA and $I$=5.28mA, respectively, $B$=400mT and $D$=1.5mJ/m$^2$). The cylindrical profile of the spin-wave radiation evolves into a spiral-like profile for 1.5mJ/m$^2$<$D$<3.0mJ/m$^2$, and a sufficiently large $I$ (SpM region) (see as an example Supplementary Movie 5). The identification of the scenario leading to the radiation of these *spiral* spin-wave modes is one of the most important results of this study.

The SKY region is observed starting from $D$ values near the critical value of $D_C$=3.16mJ/m$^2$ ($D_C = 4\sqrt{A(K_u + (B_Z - 0.5\mu_0 M_S)M_S)}/\pi$)[32]. The skyrmion nucleation process, driven by the SHE, occurs together with the excitation of propagating spin-waves (see for an example Supplementary Movie 6), and at the current $I_{sky}$ (solid line between the point 'A' and 'B'). The $I_{sky}$ curve coincides with $I_{th}$ for $D$ values larger than 4mJ/m$^2$ (point 'A' in Fig.1c). This fact constitutes the second key result of this study, i.e. the prediction that a pure spin-current with in-plane polarization can be used for the nucleation of skyrmions. The regions US/SKY, SLM/SKY and SLM/SpM are the bistability regions. In the first of these regions, we have a uniform ground state or skyrmions, depending on the excitation history. The second region is characterized by the excitation of SLMs that can coexist with the presence of skyrmions after they have been nucleated at the current $I_{sky}$. In the last region, an SLM (SpM) is observed, if the current is increased (decreased) from SLM (SpM) region. The origin of this hysteretic behavior will be discussed in detail below. We have also investigated the role of the Oersted field, finding that it does not influence qualitatively the results of Fig.1c (see Supplementary Note 2 for more details).

**Excitation of Slonczewski linear spin-wave mode.** As it was pointed out earlier, the *i*-DMI leads to the excitation of nonreciprocal spin-waves. It can be observed qualitatively in the Supplementary Movies 3 and 4, and by comparing the mode profile in Figs.2a and 2b. The largest nonreciprocal effect induced by the *i*-DMI occurs in the direction perpendicular to the in-plane projection of the static magnetization $M_0$ (*x*-axis), while the propagation along the in-plane projection of $M_0$ (*y*-axis) is reciprocal, and is characterized by the wave number that is the same as in the case of zero *D* (0.03 and 0.035nm$^{-1}$ at *I*=4.22mA and *I*=5.28mA, respectively, see Supplementary Note 2). Those results are consistent with the previous experimental measurements[33] and the results of the analytical theory[12]. The *i*-DMI-induced appearance of the nonreciprocal spin-waves leads to the decrease of the threshold current (Fig.1d), and to a "red" shift of the generation frequency at a constant current (Fig.2c). Fig.2d summarizes the dependence of the wave numbers ($|k_{-x}|$ and $|k_{+x}|$) on *D* computed from the spatial distribution of the magnetization for *I*=4.22mA and *I*=5.28mA. The difference between the $|k_{-x}|$ and $|k_{+x}|$ is shown in Fig.2e, and, as it can be noticed, is independent of *I* and *D*. All these numerically obtained features can be understood using a simple one-dimensional analytical model. In the framework of this model, we consider only the spin-waves propagating along the *x*-direction, where the spin-waves exhibit the largest nonreciprocity.

The frequency and wave vectors of the excited spin-waves are defined by the spatial quantization rule, which is determined by the spatial distribution of the spin-current $J_s$. In the case of a nonreciprocal spectrum, the general quantization rule can be written as $f(k_+ - k_-, J_s(x)) = 0$ [34], or, equivalently $k_+ - k_- = const = f_1(J_s(x))$, where $k_+$ and $k_-$ are the wave vectors of spin-waves propagating in opposite directions along the direction of maximum nonreciprocity and having the same frequency (for a reciprocal wave spectrum this rule is reduced to the condition $|k| = const$). The approximate spin-wave spectrum in the *x*-direction can be written as $\omega_k \approx \omega_0 + \omega_M \tilde{\lambda}^2 k_x^2 - \omega_M \tilde{D} k_x$ (see Methods), where $\omega_0$ is the angular frequency of the ferromagnetic resonance in the SHO, $\tilde{D} = 2D/(\mu_0 M_S^2) \sin \theta_M$, $\tilde{\lambda}^2 = \lambda^2 (2\omega_H + \omega_M (1 - H_{an}/M_S) \sin^2 \theta_M)/2\omega_0$, $\lambda$ is the exchange length in the material of the SHO ferromagnetic layer, and $H_{an} = 2K_u/(\mu_0 M_S)$ is the anisotropy field. From this equation, the wave vectors of counter-propagating nonreciprocal spin-waves, having the same frequency $\omega$, can be computed as:

$$k_{\pm x} = \frac{1}{2\omega_M \tilde{\lambda}^2} \left( \omega_M \tilde{D} \mp \sqrt{\omega_M^2 \tilde{D}^2 + 4\omega_M \tilde{\lambda}^2 (\omega - \omega_0)} \right) \quad (1)$$

where $k_{-x}$ ($k_{+x}$) is associated with the plus (minus) sign in the second term in the circular brackets in the equation (1). Substituting the wavenumber of equation (1) in the quantization rule, we get the condition $\sqrt{\omega_M^2 \tilde{D}^2 + 4\omega_M \tilde{\lambda}^2 (\omega - \omega_0)} = const_1$ that gives the following dependence of the generation frequency on *D*: $\Delta \omega = -\omega_M \tilde{D}^2/(4\tilde{\lambda}^2)$. Thus, the generation frequency has a "red" shift with the increased *D*, as obtained from our micromagnetic simulations (see Fig.2c). This effect could be easily understood by noting that the minimum spin-wave frequency in the spectrum becomes lower with the increase of *D*. From equation (1), it is easy to calculate the difference between the wave numbers of the excited waves:

$$k_{-x} - k_{+x} = \tilde{D}/\tilde{\lambda}^2 \qquad (2)$$

and to verify that this difference is independent of the quantization constant and, therefore, of the spatial distribution of the spin-current. Hence, the condition (2) can be used for the experimental determination of the magnitude and sign of the *i*-DMI parameter. Equation (2) gives a reasonable description of the simulation data, considering the same physical parameters of the SHO (Fig.2e). The fact that the dependence $\Delta k(D)$ is almost the same for different *I* is linked to a weak nonlinear variation of the spin-wave spectrum with driving current, due to a small difference in amplitudes of the excited spin-waves. Therefore, the difference of the spin-wave numbers is mainly determined by the linear spin-wave spectrum. From an experimental point of view, a direct determination of *D* can be achieved by measuring the wavelength of the emitted spin-waves along the +*x* and −*x* direction, using the phase-resolved micro-focused Brillouin light scattering[35] or time-resolved Kerr microscopy[36]. However, this method of determination of *D* may have practical limitations due to the fact that the wavelength of the excited spin-waves (see Fig.2d) are in the range 0.13-0.63μm, i.e. being comparable with the lateral resolution of the above mentioned optical techniques.

Within the above described one-dimensional model, we can also calculate the dependence of the threshold current for spin-wave excitation on *D*. Assuming a rectangular profile of the charge current density in the active region ($J(x) = J$ within $x = [0, d]$, and $J(x) = 0$ otherwise), one can get the following implicit expression (similar to equation (6c) in [37]),

$$\left(\bar{k} + i\frac{\Gamma_G - \Gamma_J}{v}\right)\tan\left[\left(\bar{k} + i\frac{\Gamma_G - \Gamma_J}{v}\right)\frac{d}{2}\right] = -i\left(\bar{k} + i\frac{\Gamma_G}{v}\right) \qquad (3)$$

where $\bar{k} = (k_{-x} - k_{+x})/2$ is the average wave number of excited nonreciprocal spin-waves (note, that in our notation $k_{-x} < 0$), *d* is the distance between the SHO golden electrodes characterizing the spatial localization of the spin-current, $v = \left[\omega_M^2 \tilde{D}^2 + 4\omega_M(\omega - \omega_0)\tilde{\lambda}^2\right]^{(-1/2)}$ is the spin-wave group velocity, $\Gamma_G = \alpha_G \omega$ is the spin-wave damping, $\Gamma_J = \sigma J$ is the negative damping created by the spin-current, and $\sigma = g\mu_B \alpha_H \sin\theta_M /(2eM_s t_{CoFe})$ determines the spin-Hall efficiency (*g* is the Landè factor, $\mu_B$ the Bohr magneton, *e* the electronic charge and $t_{CoFe}$ the CoFe layer thickness). The threshold current calculated from equation (3) is compared with numerical results in Fig.1d by using the value $I_{th}$=3.7mA at *D*=0mJ/m² as a fitting parameter. One can see a good coincidence between the analytical and numerical results. Note that the decrease of the threshold current with *D* has the same nature, as a frequency "red" shift-lowering the bottom of the spin-wave spectrum with the increase of $D$ and, consequently, the decrease of spin-wave damping $\Gamma_G = \alpha_G \omega$.

The SLM in SHOs have not been observed experimentally, since the threshold current for their excitation is expected to be very large (>$10^9$A/cm$^2$)[20], around three times larger than the current necessary to excite a "bullet" spin-wave mode in an SHO with in-plane magnetization. In the SHO of this study, we were able to reduce the critical current density of one order of magnitude (<$4\times10^8$A/cm$^2$) thanks to the additional perpendicular interface anisotropy in the CoFe ferromagnet. This additional anisotropy allows one to achieve the positive nonlinear frequency shift, required for the SLM excitation[31], at a higher magnetization angle $\theta_M$, which results in the higher spin-Hall efficiency, since it is proportional to $\sin\theta_M$. A further reduction of the current density can be achieved by including an additional Ta layer above the CoFe ferromagnet[24].

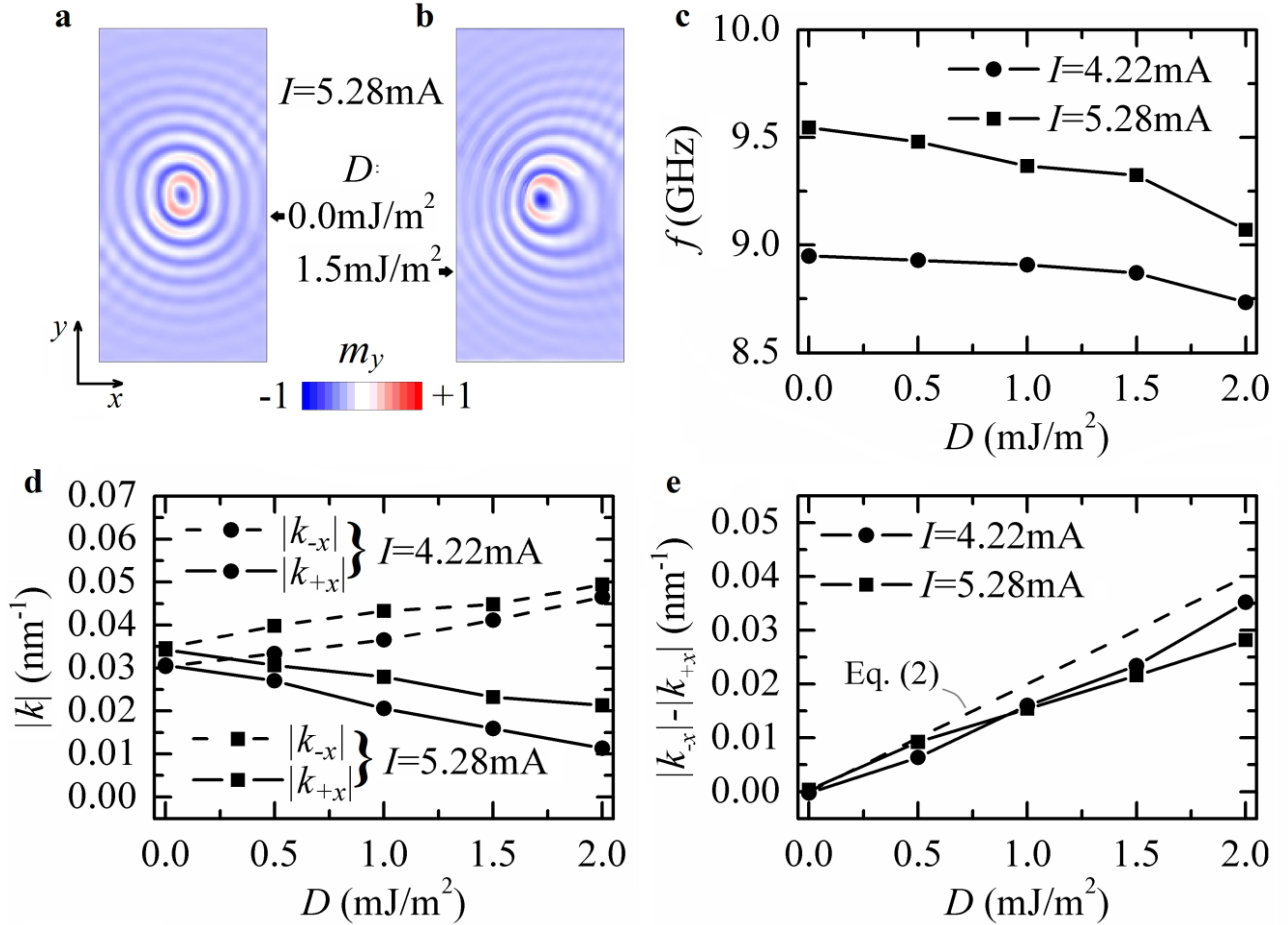

**Figure 2. | Non-reciprocal Slonczewski linear modes. a**, Oscillation frequency of the excited SLM as a function of $D$ for two values of the driving current. **b** and **c**, Example of the spatial profile of the reciprocal and nonreciprocal Slonczweski spin waves, respectively, calculated for $I$=5.28 A and $D$=0 and 1.5mJ/m$^2$, respectively. **d**, Wave numbers along the -$x$ and +$x$ directions as functions of $D$ for two values of current; **e**, Difference between the wave numbers along the positive and negative $x$-directions as a function $D$ for the same two values of current (solid lines) and the same difference calculated analytically from equation (2) (dashed line).

**Excitation of spin-wave modes with a spiral spatial profile.** Fig.3a summarizes the spin-wave frequency as a function of *I* computed for $D$=0.0mJ/m$^2$ and $D$=1.5mJ/m$^2$ (*d*=100nm). In the absence of the *i*-DMI, the oscillation frequency shows a monotonic increase with current, or a "blue" frequency shift, typical for the Slonczewski linear propagating spin-wave mode. A different frequency behavior is seen for $D$=1.5mJ/m$^2$, where the frequency tunability with current becomes non-monotonic. This behavior is robust under the variation of *d*, as seen from Fig.3b where *d*=200nm. At sufficiently large *I* and *D*, the spin-wave is converted from the cylindrical to a spiral-like (SpM region in Fig.1c). Fig.3c shows a spiral-type profile (the color is linked to the *y*-component of the magnetization).

In order to understand the origin of the spiral mode, we have performed a detailed analysis of the spatial distribution of the dynamic magnetization in the SHO ferromagnetic layer in this regime. Figs.3d-g illustrate four snapshots (*I*=6.33mA) which clearly reveal the physics of the spiral mode formation. In the SpM region, the SHE is able to nucleate a dynamical soliton[38,39,40]. It is characterized by a central core with the magnetization pointing along the negative out-of-plane direction (opposite to the equilibrium axis of the magnetization), and by the rotation of its boundary spins through 360° (see Figs.3d-g). The dynamical skyrmion exhibits a rotational motion (gyration) along a circular trajectory within the region of the high current density, that is typical for solitons with nonzero topological charge under the influence of spin-current[41] (see Supplementary Movie 7). Dynamical skyrmion plays a role of a "source" for magnetization oscillations in the outer region, and, since the source is gyrating, the radiation acquires the form of a spiral wave, as it happens in many other fields with gyrating source[42,43]. Note, also, that once it has been excited the SpM is still stable at lower current magnitudes in the SpM/SLM region, because the excitation of the dynamical skyrmion is linked to a sub-critical Hopf bifurcation[40].

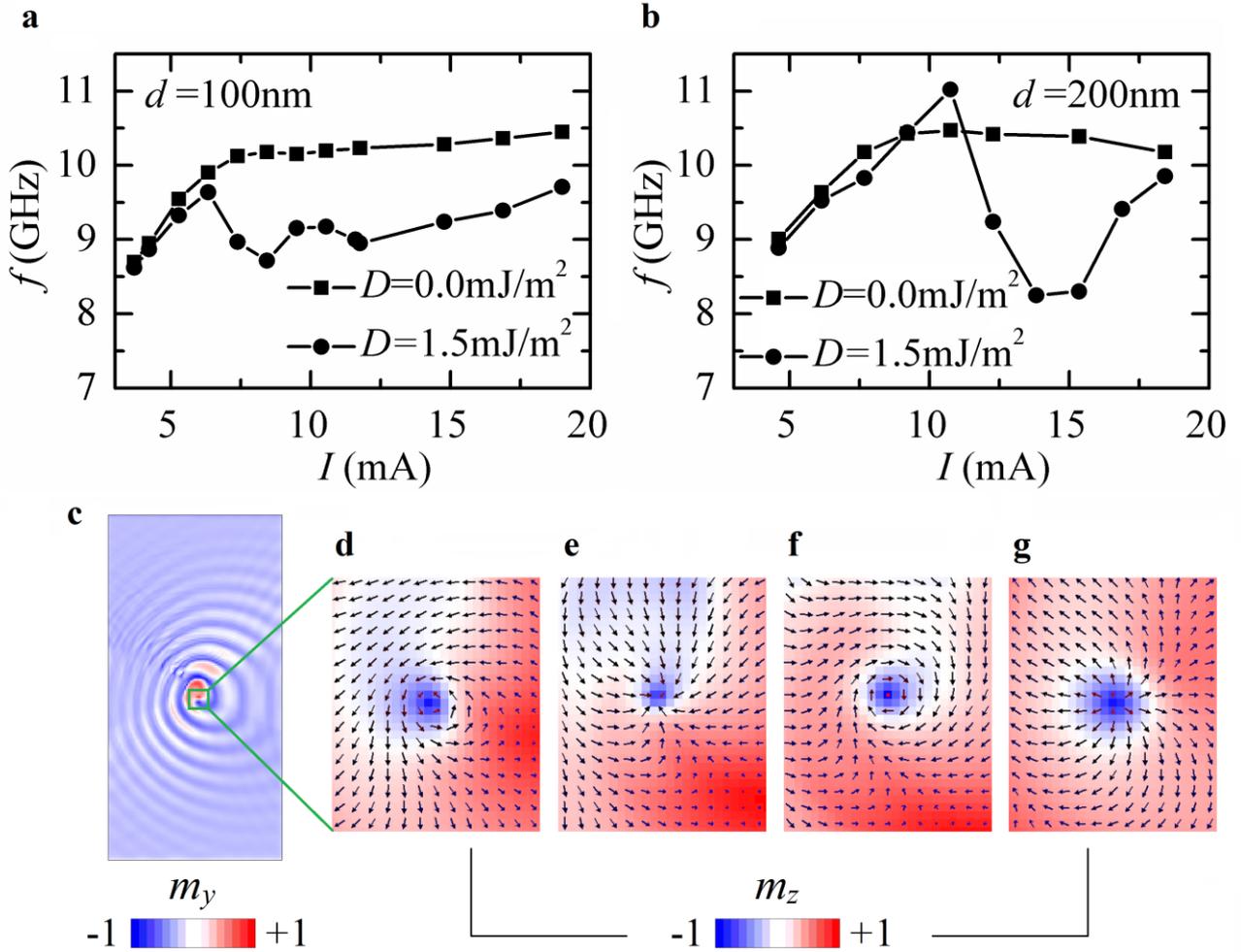

**Figure 3 | Excitation of a spin-wave with spiral profile. a**, Oscillation frequency of the excited mode as a function of $I$ without and with $i$-DMI, for $d$=100 nm. **b**, Same as **a**, but with $d$=200 nm. **c**, Example of a spatial profile of the spiral-type spin-wave for $I$=6.22mA and $D$=1.5mJ/m². **d-g**, Spatial distributions of the magnetization characterizing a topological-type magnetic soliton, the current-induced gyration of which causes the radiation of a spiral-type spin wave mode.

**Generation of single skyrmions and "gas" of skyrmions.** The last regions of the phase diagram of Fig.1c are related to skyrmions. For the critical $D_C$, the skyrmions become energetically stable[32] and, after the nucleation driven by the SHE (SKY region) (see Supplementary Movie 6 for the nucleation of a single skyrmion), they remain stable even when the driving current is switched off (US/SKY region). Once the skyrmion is nucleated, it is shifted along the spin-current direction, as expected for Néel skyrmions[16]. For $D$ below 4.0mJ/m² (point 'A' in Fig.1c), $I_{sky}$ and $I_{th}$ split into different curves, and, hence, in the SKY/SLM region when the current increases from the uniform state, only the SLMs are excited. The presence of this region in the phase diagram is interesting from a fundamental point of view, as it identifies a scenario where the interaction between the spin-waves and skyrmions[44] can be studied. Fig.4a shows the nucleation time of a single skyrmion as a function of the current magnitude for two values of $D$ (3.5 and 4.0mJ/m$^2$). It can be seen from Fig.4a that a sub-nanosecond skyrmion nucleation time can be achieved (see Fig.4b for a single skyrmion snapshot). Our results predict a new scenario for a single skyrmion nucleation driven by a pure spin-current. This method can be used as an alternative to the method based on the STT from a perpendicular spin-polarized current[15], with the possible advantage of the simpler fabrication process of the device. If the current is not switched off, the skyrmions are nucleated continuously. However, since the current is non-uniformly applied, the skyrmions tend to accumulate in one side of the ferromagnet. A skyrmion "gas" is, therefore, formed[45] (see Fig.4c for an example of the spatial distribution of the skyrmions). When the "gas" saturates, i.e. when no more skyrmions can be nucleated in the "gas" because of the skyrmion-skyrmion magnetostatic repulsion, each skyrmion further nucleated is immediately annihilated (see Supplementary Movie 8). This result paves the way to study the magnetic properties of skyrmion "gas" described theoretically in[45].

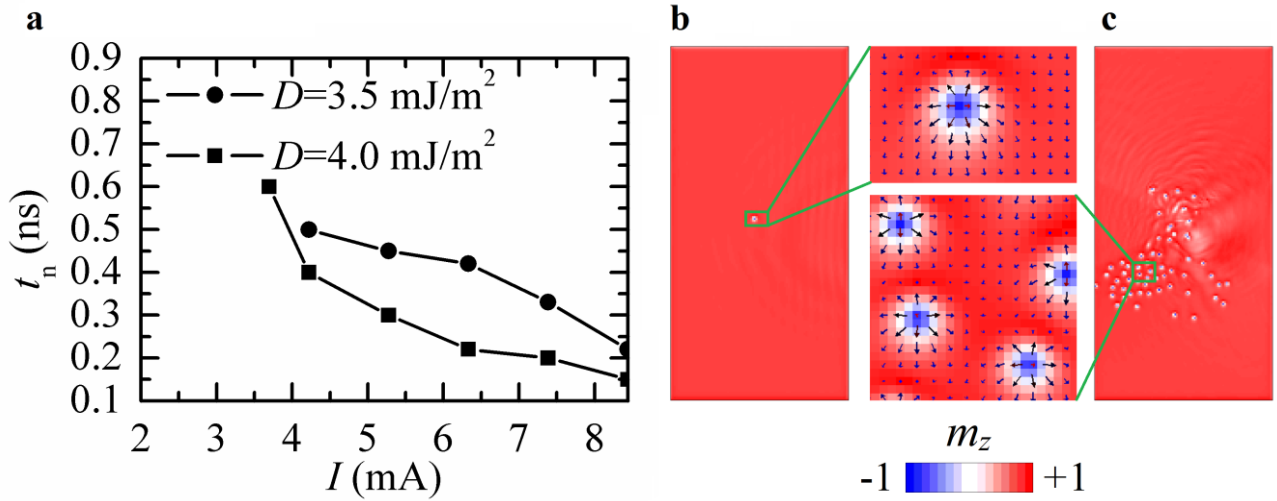

**Figure 4 | Skyrmion nucleation. a**, Nucleation time of a single skyrmion as a function of the current amplitude for $D$ = 3.5 and 4.0mJ/m$^2$. **b**, Snapshot of a single skyrmion and zoom of the skyrmion nucleation region. **c**, Snapshot of a skyrmion gas and zoom of the region indicated by a green square in the right frame.

## Discussion

In our study, we propose an SHO device geometry that, combining SHE and $i$-DMI, offers a unique opportunity to study nonreciprocal effects of spin-wave propagation in two dimensional systems and to observe a new type of dynamical spin-wave modes having a spiral spatial profile. This novel spin-wave mode originates from the gyrotropic rotation of a dynamical skyrmion. From the technological point of view, the proposed SHO geometry could be useful for the development of novel generators of short propagating spin-waves in future magnonic signal processing devices. From the fundamental point of view, it is also very interesting, as it allows to study the interaction of spin-wave and skyrmions, as well as to control the number of the nucleated skyrmions by applying a properly designed current pulse.

**Methods**

**Micromagnetic framework.** Micromagnetic simulations were carried out by means of a *state-of-the-art* parallel micromagnetic solver, which numerically integrates the LLG equation including the Slonczewski-like torque due to SHE[46,20]:

$$\frac{d\mathbf{m}}{d\tau} = -\mathbf{m} \times \mathbf{h}_{EFF} + \alpha_G \mathbf{m} \times \frac{d\mathbf{m}}{d\tau} - \frac{g\mu_B}{2\gamma_0 e M_S^2 t_{CoFe}} \alpha_H \mathbf{m} \times \mathbf{m} \times (\hat{z} \times \mathbf{J}) \quad (4)$$

where $\mathbf{m}$ and $\mathbf{h}_{EFF}$ are the normalized magnetization and the effective field of the ferromagnet. The effective field includes the standard magnetic field contributions, as well as the *i*-DMI and Oersted field (see also Supplementary Note 1). $\tau$ is the dimensionless time $\tau = \gamma_0 M_S t$, where $\gamma_0$ is the gyromagnetic ratio, and $M_s$ is the saturation magnetization of the ferromagnet. $\alpha_G$ is the Gilbert damping, $g$ is the Landè factor, $\mu_B$ is the Bohr Magneton, $e$ is the electron charge, $t_{CoFe}$ is the thickness of the ferromagnetic layer, $\alpha_H$ is the spin-Hall angle obtained from the ratio between the spin current and the electrical current. $\hat{z}$ is the unit vector of the out-of-plane direction and $\mathbf{J}$ is the in-plane current density injected via the heavy metal. The *i*-DMI energetic density expression, as derived considering the ultra-thin film hypothesis $\left(\frac{\partial \mathbf{m}}{\partial z} = 0\right)$, is $\varepsilon_{i-DMI} = D\left[m_z \nabla \cdot \mathbf{m} - (\mathbf{m} \cdot \nabla)m_z\right]$ $D$ being the parameter taking into account the intensity of the DMI, and $m_z$ is the *z*-component of the normalized magnetization. By making the functional derivative of equation, the normalized *i*-DMI effective field is given by:

$$\mathbf{h}_{i-DMI} = -\frac{1}{\mu_0 M_S^2} \frac{\delta \varepsilon_{i-DMI}}{\delta \mathbf{m}} = -\frac{2D}{\mu_0 M_S^2}\left[(\nabla \cdot \mathbf{m})\hat{z} - \nabla m_z\right] \quad (5)$$

The boundary conditions related to the interfacial DMI are expressed by $\frac{d\mathbf{m}}{dn} = \frac{1}{\xi}(\hat{z} \times \mathbf{n}) \times \mathbf{m}$ where $\mathbf{n}$ is the unit vector normal to the edge and $\xi = \frac{2A}{D}$ (being $A$ the exchange constant) is a characteristic length in the presence of *i*-DMI.

We have studied a bilayer system Pt(5nm)/CoFe(1nm) with a rectangular cross section of 1500x3000nm$^2$. The electric current was locally injected into the ferromagnet via a thick Au electrode (thickness of 150nm) with two tips located at a distance $d$ from each other. The charge current flowing in the Pt layer gives rise to the SHE and then, to flow of perpendicular (along the "z" axis) pure spin current at the Pt/CoFe interface creating an anti-damping Slonczewski-like torque in the ferromagnetic layer. At sufficiently large magnitudes of the charge current his torque compensates the Gilbert losses in the ferromagnetic layer and excites in it persistent magnetization oscillations. For the results discussed in the main text, we have considered the following physical parameters of the SHO (Fig.1a): saturation magnetization $M_S=1\times10^6$ A/m[47], exchange stiffness constant $A=2.0\times10^{-11}$ J/m, interfacial perpendicular anisotropy induced at the boundary between CoFe and Pt characterized by the anisotropy constant $K_u=5.5\times10^5$ J/m$^3$,[48] damping constant $\alpha_G=0.03$[49], and the spin-Hall angle $\alpha_H=0.1$[9]. The ferromagnetic CoFe layer has an in-plane equilibrium magnetization at zero field which is directed along the $y$- in-plane direction due to the shape anisotropy of the ferromagnetic layer. The real spatial distributions of the density $J_e$ of the charge current, density $J_s$ of the spin current and the Oersted field were calculated numerically, as it is described in the Supplementary Note 1.

**Derivation of analytical equations.** The spin wave dispersion relation along the $x$-direction in the presence of $i$-DMI can be calculated analogously to Ref. [12] and has the following form

$$\omega_k = \sqrt{\left(\omega_H + \omega_M \lambda^2 k_x^2\right)\left(\omega_H + \omega_M \lambda^2 k_x^2 + \omega_M\left(1 - H_{an}/M_S\right)\sin^2\theta_M\right)} - \omega_M \tilde{D} k_x, \qquad (6)$$

where $\omega_H = \gamma B_{eff}$, $H_{eff}$ is the static effective field, $\omega_M = \gamma \mu_0 M_s$. In the range $\omega_M \lambda^2 k^2 \ll \omega_0$ it can be approximated as $\omega_k \approx \omega_0 + \omega_M \tilde{\lambda}^2 k_x^2 - \omega_M \tilde{D} k_x$, where $\omega_0 = \sqrt{\omega_H \left(\omega_H + \omega_M\left(1 - H_{an}/M_s\right)\sin^2\theta_M\right)}$ is the angular frequency of the ferromagnetic resonance in the ferromagnetic layer, and $\tilde{\lambda}^2 = \lambda^2(2\omega_H + \omega_M(1-H_{an}/M_S)\sin^2\theta_M)/2\omega_0$.

Making a formal substitution $k_x \to -i(d/dx)$ in this dispersion equation, it is possible to obtain the following dynamical equation describing the spatial and temporal evolution of the spin wave complex amplitude $a$:

$$\frac{\partial a}{\partial t} = -i\omega a = -i\left(\omega_0 - \omega_M \tilde{\lambda}^2 \frac{\partial^2}{\partial x^2} + i\omega_M \tilde{D}\frac{\partial}{\partial x}\right)a - \alpha_G \omega a + \sigma J(x)a. \qquad (7)$$

The spin wave damping is accounted for by the term $\alpha_G \omega$ (spin wave ellipticity, which could modify the damping term[50] in our case is small), while the influence of the spin current could be easily calculated from equation (4) within the framework of the perturbation theory[50] and is given by the term $\sigma J(x)a$, with $\sigma = g\mu_B \alpha_H \sin\theta_M / (2eM_S t_{CoFe})$. Equation (3) for the threshold current can be obtain, analogously to Ref[13], by deriving general analytical solutions of equation (7) inside and outside the current-carrying region, and applying the boundary conditions of continuity for the spin wave complex amplitude *a* and its derivative. It is clear, that in the reciprocal case, when $k_{-x} = -k_{+x}$, equation (3) is reduced to equation (6) from Ref.[37].

## Acknowledgments

This work was supported by the project PRIN2010ECA8P3 from Italian MIUR, the bilateral agreement Italy-Turkey (TUBITAK-CNR) project (CNR Grant #: B52I14002910005, TUBITAK Grant #113F378) "Nanoscale magnetic devices based on the coupling of Spintronics and Spinorbitronics", and the executive programme of scientific and technological cooperation between Italy and China for the years 2016-2018 (code CN16GR09) title "Nanoscale broadband spin-transfer-torque microwave detector" funded by Ministero degli Affari Esteri e della Cooperazione Internazionale. A.S and R.V. acknowledge support from the Grant ECCS-1305586 from the National Science Foundation of the USA, from the contract with the US Army TARDEC, RDECOM, from DARPA MTO/MESO grant N66001-11-1-54114, and from the Center for NanoFerroic Devices (CNFD) and the Nanoelectronics Research Initiative (NRI). The authors thank Domenico Romolo for the graphical support. R. T. acknowledges Fondazione Carit - Projects – "Sistemi Phased-Array Ultrasonori", and "Sensori Spintronici".

## Author contributions

A.G., R.Z., M.C., G. G. and G.F. initiated the work and designed the numerical experiments. R.V. and A.S. developed the analytical theory and performed the analytical calculations. A.L. performed the computation of the spatial distribution of the current density and the Oersted field and wrote the supplementary note 1. A.G. performed micromagnetic simulations supported by V.P., G.S., B.A and M.C. V.P. wrote the supplementary note 2 and prepared the last version of the figures. A.G., and G.F. analyzed the data. G.F. wrote the paper with input from R.V., A.S. and R.T. All authors contributed to the general discussion of the results and commented on the manuscript.

## Additional information

**Supplementary Information** accompanies this paper at ...

**Competing financial interests**: The authors declare no competing financial interests.

**Reprints and permission information** is available online at ...

**How to cite this article**: Xxxx